\documentclass[times]{akari2017} 

\usepackage{here}

\shorttitle{Microwave dust emission and AKARI}

\begin{document}

\title{A look at possible microwave dust emission via AKARI infrared all-sky surveys}



\correspondingauthor{Aaron C. Bell}
\email{abell@astron.s.u-tokyo.ac.jp}

\author{Aaron C. Bell}
\affiliation{Graduate School of Science, The University of Tokyo, Bunkyo-ku, 113-0033, Tokyo, Japan}

\author{Takashi Onaka}
\affiliation{Graduate School of Science, The University of Tokyo, Bunkyo-ku, 113-0033, Tokyo, Japan}

\author{Yasuo Doi}
\affiliation{Department of Earth Science and Astronomy, The University of Tokyo, 3-8-1 Komaba, Meguro-ku, Tokyo 153-8902, Japan}

\author{Fr\'ed\'eric Galliano}
\affiliation{Service d'Astrophysique, CEA/Saclay, L'Orme des Merisiers, 91191 Gif-sur-Yvette, France}

\author{Ronin Wu}
\affiliation{LERMA, Observatoire de Paris, PSL Research University, CNRS, Sorbonne Université, UPMC Paris 06, 92190, Meudon, France}

\author{Hidehiro Kaneda}
\affiliation{Nagoya University, Furo-cho, Chikusa-ku, Nagoya 464-8602, Japan}

\author{Daisuke Ishihara}
\affiliation{Nagoya University, Furo-cho, Chikusa-ku, Nagoya 464-8602, Japan}

\author{Martin Giard}
\affiliation{Université de Toulouse, UPS-CNRS, IRAP, 31028, Toulouse Cedex 4, France}



\begin{abstract}

The anomalous microwave emission (AME) still lacks a conclusive explanation.  This excess of emission, roughly between 10 and 50~GHz, correlates spatially with interstellar dust, prompting a ``spinning dust'' hypothesis:  electric dipole emission by rapidly rotating, small dust grains. The typical peak frequency range of the AME profile implicates grains on the order of \textasciitilde{}1nm, suggesting polycyclic aromatic hydrocarbon molecules (PAHs). We compare AKARI/Infrared Camera (IRC), with its thorough PAH-band coverage, to AME intensity estimates from the Planck Collaboration, in the $\lambda$~Orionis region. We look also at infrared dust emission from other mid IR and far-IR bands. The results and discussion contained here apply to an angular scale of approximately 1$^{\circ}$. In general, our results support an AME-from-dust hypothesis. In $\lambda$~Orionis, we find that certainly dust mass correlates with AME, and that PAH-related emission in the AKARI/IRC 9~$\mu$m band may correlate slightly more strongly.

\end{abstract}



\setcounter{page}{1}



\section{Introduction} \label{sec:intro}

  In our efforts to decompose and understand galactic microwave emission itself, there remains a constant antagonist. Galactic foregrounds had been broken down into three dominant components: free-free emission from ionized regions, synchrotron emission generated by electrons relativistically by the Milky Way's magnetic field, and the microwave extent of thermal dust emission \citep{planckXII}. Deviations from this understanding began to appear in the early 1990s, with efforts by \cite{kogut96, leitch97} to carefully investigate the CMB. They had found a component of the microwave sky which implied unlikely spectral indices for free-free or synchrotron emission. This ``anomalous microwave emission'' (AME) generally takes the form of an 'excess' continnuum emission source, having a peak somewhere between 10 to 40~GHz. This excess defies predictions for known microwave emission mechansisms. AME still lacks a concrete physical explanation.

  AME has been found to be a widespread feature of the microwave Milky Way. \cite{kogut96,deoliveiracosta97} showed that the AME correlates very well with infrared emission from dust, via COBE/DIRBE and IRAS far-IR maps. More recent works, employing the latest IR to microwave all-sky maps, and various ground based radio observations have strongly confirmed a relationship between interstellar dust emission and AME \citep{ysard10a,tibbs11,hensley16}. A comprehensive state-of-play of AME research is given in \cite{dickinson18}.

   From the observed spatial correlation between AME and dust emerged two prevailing hypotheses:

  1) Electric dipole emission by spinning small dust grains or polycyclic aromatic hydrocarbons (PAHs).
  2) Magnetic dipole emission, caused by thermal fluctuations in grains with magnetic inclusions, proposed by \cite{draine99}.

  We explore the case that the AME signature arises from spinning dust emission. If the AME is carried by spinning dust, the carrier should be small enough that it can be rotationally excited to frequencies in the range of 10-40~GHz, and must have a permanent electric dipole. Assuming the rotational emission model of \cite{draine98b}, the AME signature (consistent with peaked, continuum emission having a peak between 15 and 50~GHz ) implies very small oscillators (\textasciitilde{}1~nm). Polycyclic aromatic hydrocarbon family of molecules (PAHs), or nanoscale amorphous carbon dust fit the size criteria. While \cite{hensley17a} claim that AME might also be explained by spinning nanosilicates, only PAHs show both:
   1) Evidence of abundance in the ISM at IR wavelengths, \citep{giard94,onaka00}, and
   2) A predicted range of dipole moments (on order of 1~debye), to produce the observed AME signature \citep{draine98b}.

  The $\lambda$~Orionis molecular ring, also known as the Meissa Ring, has been strongly highlighted by \cite{planck15XXV} for significant AME, and a prime target for testing the spinning PAH hypothesis. The ring contains an HII region, ionized by $\lambda$Ori itself and its OB associates \citep{ochsendorf15}. At approx. 10$^{\circ}$ wide, we can see the outline of the structure even in the low (1$^{\circ}$ FWHM) resolution PCAME map.

\section{Analysis} \label{sec:analysis}
\begin{figure}[!ht]
      \centering
      \includegraphics[width=0.3\textwidth]{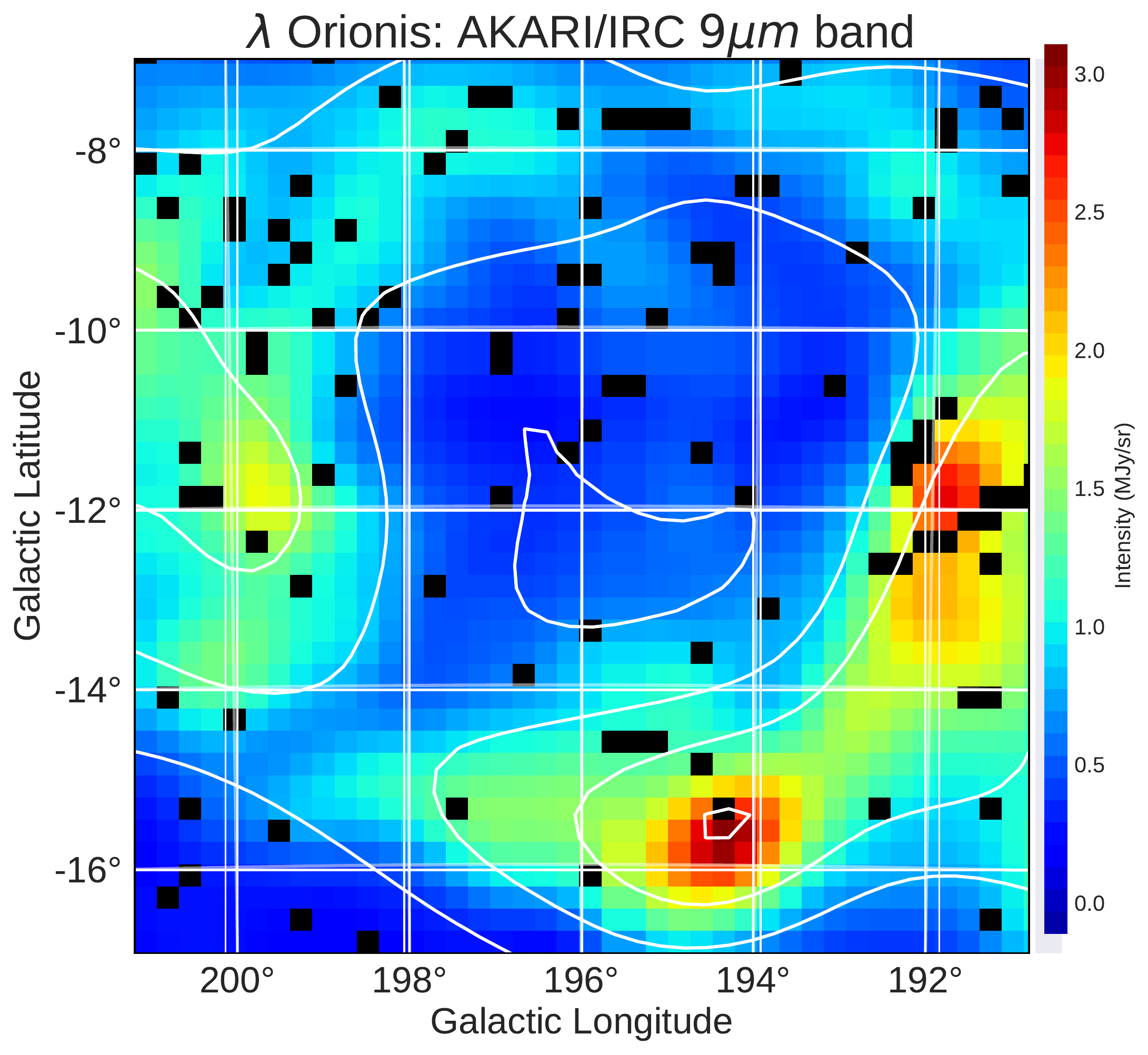}
      \includegraphics[width=0.3\textwidth,trim={2.5cm 2cm 3.0cm 2cm},clip]{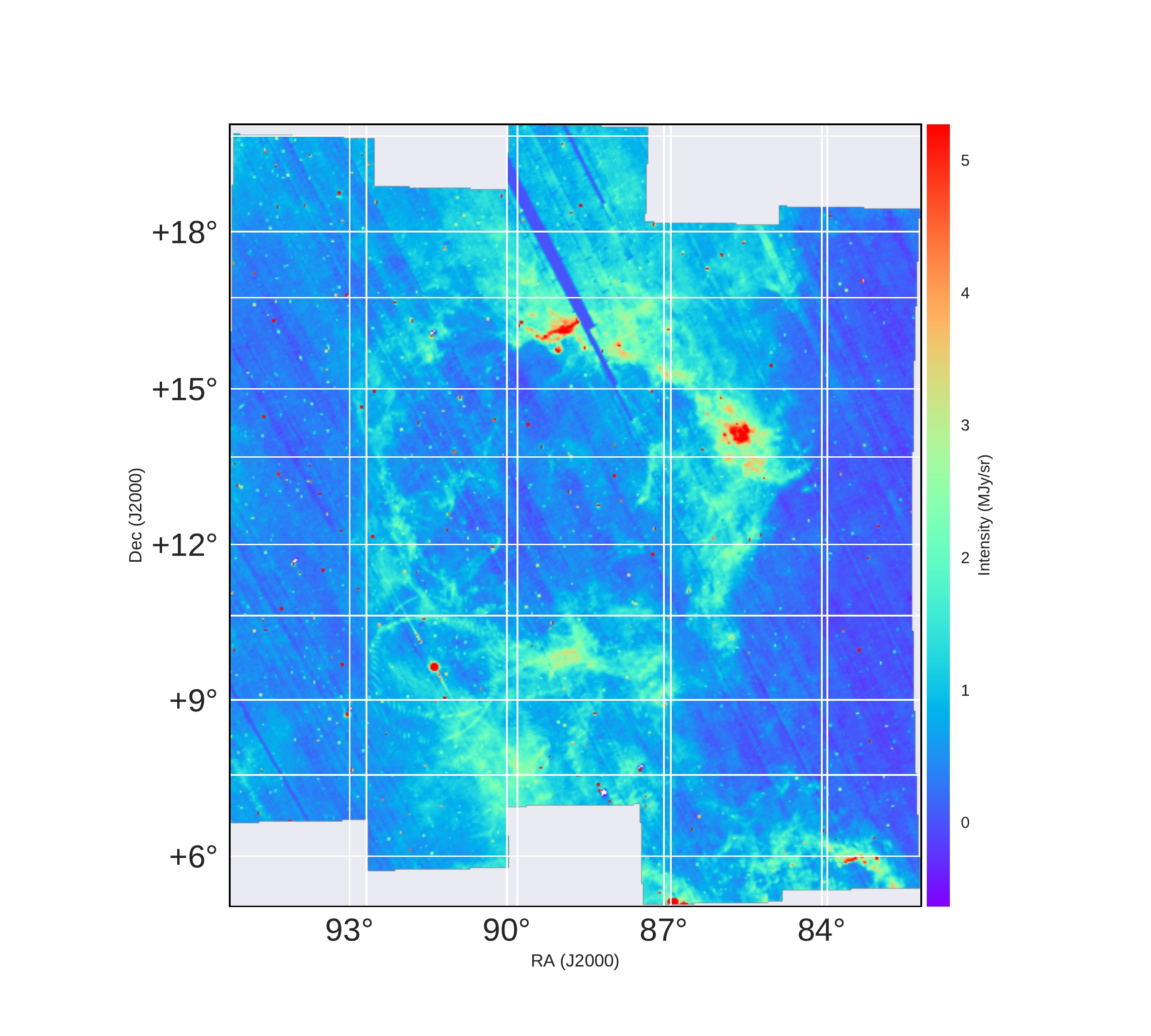}
      \caption{Left: $\lambda$~Orionis as it appears in the AKARI 9~$\mu$m data. Contours indicate the AME, as given by the Planck PR2 AME map. The image is smoothed to a 1$^{\circ}$ PSF (much larger than the original 10$''$). The $\lambda$~Orionis star itself is approximately located at the center of the image. Right: The $\lambda$~Orionis region in the A9 band at near-native resolution. This is a mosaic created from the 3x3 degree all-sky survey tiles by Ishirara et al. (in prep.) Betelgeuse, in the lower left of the image, is bright enough in this band to produce a ring-shaped artifact.}
      \label{fig:orionis-akari9}
    \end{figure}

  We have carried out an initial comparison of the AME of$\lambda$~Orionis in mid to far-IR dust emission. The region is shown in Fig.~\ref{fig:orionis-akari9} as it appears in 1$^{\circ}$-smoothed A9 data.
   The ring structure itself indicates excess microwave emission attributed to AME. We use 12 photometric all-sky maps. This includes the two IRC bands (A9 and A18) as well as MIR to FIR data from IRAS (I12, I25, I60 and I100), Planck (P545 and P857), and AKARI/FIS (A65, A90, A140, A160). For the IRC data, we produce mosiacs of $\lambda$~Orionis from the individual tiles provided in the internal all-sky archive.

  Other data are obtained via HEALPix maps. We employ the {\tt healpix2wcs} functionality provided in the {\tt gnomdrizz} python package. A9 and A18 images are produced by regridding the images with the {\tt Montage} software by NASA/IPAC. Fig.~\ref{fig:orionis-akari9} shows a high resolution mosaic of the A9 data before processing.

  We apply a pixel mask to the data, to account for outliers and systematic errors in each band. This consists of a point-source mask (mostly due to point source contaimination in the MIR bands), and a ``missing stripe error'' mask (mostly affecting the AKARI FIS data) mask. The data are smoothed to a common approximately circular Guassian PSF with FWHM 1$^{\circ}$, in order to have a resolution approximating that of the PC AME data.

\begin{figure}[!ht]
      \includegraphics[width=0.80\textwidth,trim={5cm 5cm 3.5cm 5cm},clip]{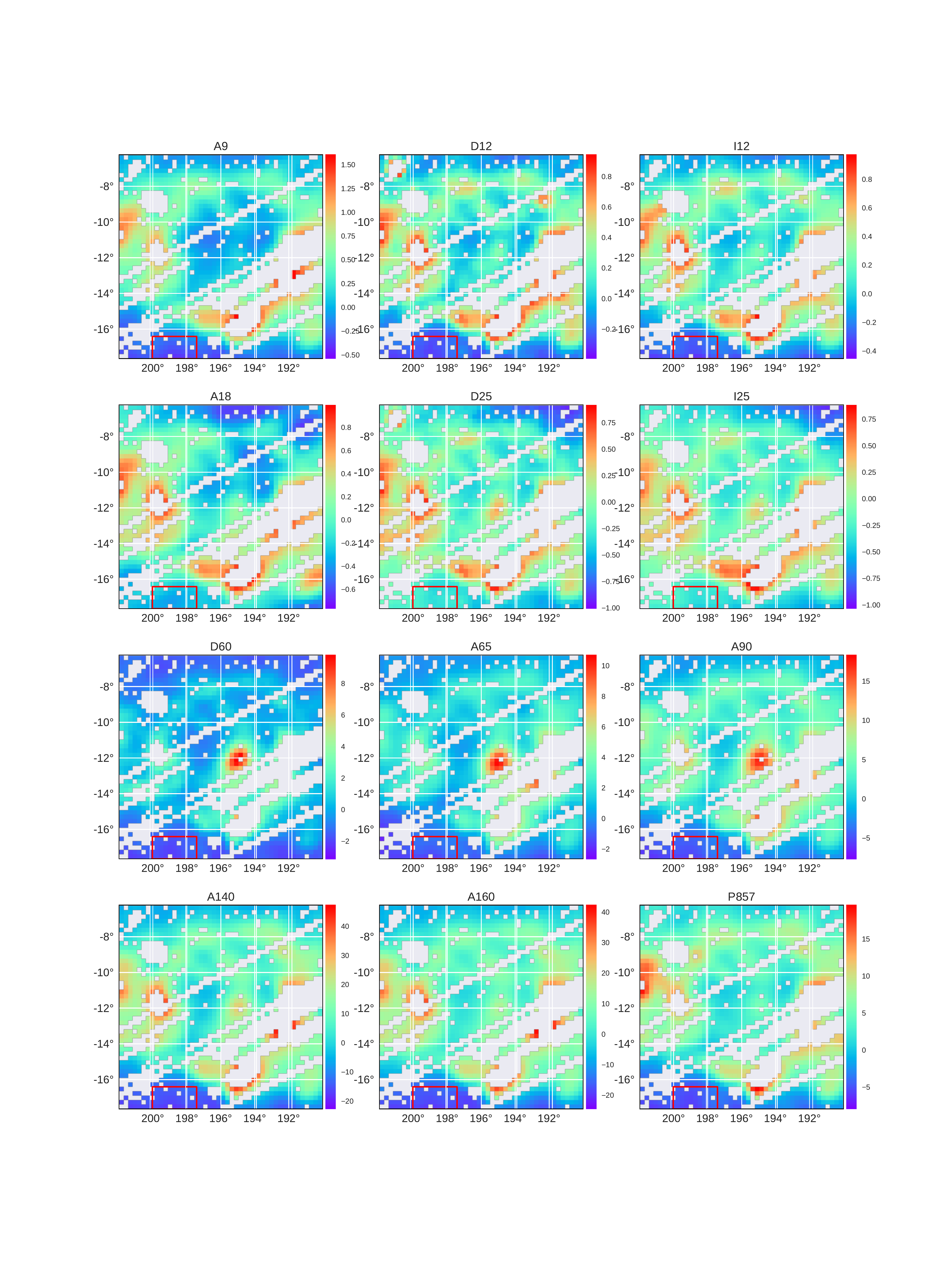}
      \centering
      \caption{Processed data at each wavelgnth for $\lambda$~Orionis. A flat background has been subtracted from each frame based on the mean of pixels within the red rectangle. The pixel width is 0.25$^{\circ}$, with the data PSF smoothed to 1$^{\circ}$ spatial resolution. Colorbars indicate the intensity in MJy/sr.}
      \label{fig:lori_processed_all}
    \end{figure}

  We estimate an average, flat background level for this region. The background level is determined the mean of pixels in an `OFF` zone. The final images are shown in Fig.~\ref{fig:lori_processed_all}, with the full mask applied (masked pixels are indicated in white), and with the OFF zone indicated by the red rectangle on each frame.

  We do not expect simple band-by-band intensity correlation tests with the AME to be sensitive to background and foreground emission along the line of sight towards the $\lambda$~Orionis region.

  We perform a Bootstrap test of the correlations between the AME and the various IR bands used in this study, to evaluate which dust emisison wavelengths correlate best with AME.

\begin{figure}[!ht]
      \includegraphics[width=0.75\textwidth,trim={3cm 0.25cm 2.5cm 1cm},clip]{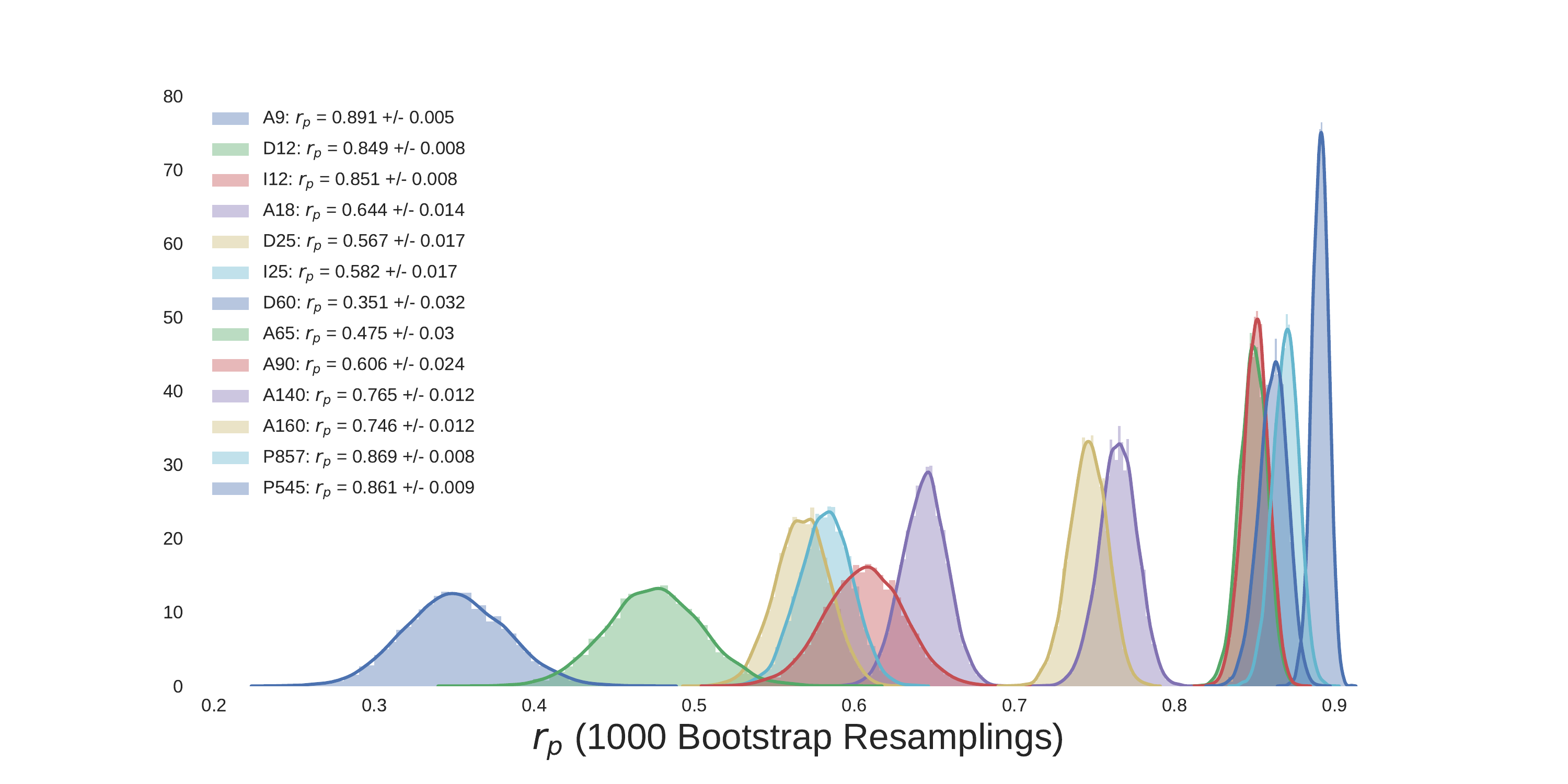}
      \centering
      \caption{Re-sampled (Bootstrap) correlation tests for IR emission in $\lambda$~Orionis vs. AME. Each band's $r_p$ distribution is shown in a different color (the same color scheme for both plots). The width of the distribution indicates the error for the given data in the correlation coefficient.}
      \label{fig:bootstrap_vs_AME}
    \end{figure}

  To assess the robustness of the correlation scores, we employ the Bootstrap re-sampling approach, first introduced by \cite{efron79}. This involves creating random re-sampled sets of the data. We use the 'with replacement' approach, meaning that a data point may be selected multiple times in a single re-sampling iteration. The size of the re-sampled set is the same is the input set size. For each random set we run a correlation test, resulting in a distribution of correlation coefficients. This provides an estimate of the error bars for the correlation scores, and serves to de-weight outliers.
  We carry out bootstrap correlation tests for each IR band's intensity vs. AME intensity. The data are resampled 1000 times for each test. The distributions of the boostrap resamplings are shown in Fig.~\ref{fig:bootstrap_vs_AME}.

\section{Results} \label{sec:results}
  This confirms a correlation between the IR and AME. Interestingly though, the correlation strengths with AME show a pattern from short to long wavelengths: A9, P857, and P545 show the strongest correlations, with the correlation weaking from A18 to A90, and again strengthening at longer wavelengths. The overall pattern is for bands dominated by PAH emission, and those which trace Rayleigh-Jeans thermal dust emission are equally good predictors of the AME. Bands dominated by a mixture of very small grains (VSGs), and warm dust emission, show a weaker correlation.

  Comparing the images in Fig.~\ref{fig:lori_processed_all}, most of the variation in the correlation scores appears to come from the central region of $\lambda$~Orionis. Because of the known heating present within the ring, from the $\lambda$~Orionis association, and given the brightening of bands between A18 and A90, this variation appears to be due to a temperature increase.

\section{Discussion} \label{sec:discussion}
  In  $\lambda$~Orionis we found that accross the whole region, A9 emission and P545 emission were the most strongly correlated with AME. This is apparent both in the photometric band analysis, and in the dust SED fitting.  The fact that the correlation strengths of PAH-tracing mission and sub-mm emission are similar is in-line with \cite{ysard10b} and \cite{hensley16}--- although these two papers are odds as to which relationship is stronger.

   The results are consistent with a scenario in which PAH mass, cold dust, and the AME are all tightly correlated. Weaker correlation from 25 to 70~$\mu$m may indicate that AME is weaker in regions of warmer dust and stronger radiation fields. Such an anti-correlation with harsher radiation are consistent with the carriers of AME being destroyed in the central region of $\lambda$~Orionis. We cannot conclusively identify PAHs at the AME carrier, nor can we rule out nanosilicates.

  Examining $\lambda$~Orionis in intensity, we find that the A9 intensity correlates more strongly with AME than I12 or D12, the other PAH-tracing bands. In fact, A9 correlates more strongly with AME than any other band. This is consistent with the spinning PAH hypothesis, and taken alone may indicate that the 6.2~$\mu$m feature emission from charged PAHs, may be a marginally better predictor of AME intensity. This could be consistent with PAH anions surviving in the portions of $\lambda$~Orionis which are emitting the strongest AME.

  Future wide-area spectral mapping of the $\lambda$~Orionis region may be able to conclusively test for increased $fPAH+$ in regions with stronger AME. Such studies would be strongly aided by higher resolution probing of spatial variations in the AME spectral profile. The ionization fraction of PAHs also may be worth further investigation in the context of AME. \\

  This research is based on observations with AKARI, a JAXA project with the participation of ESA.

\bibliography{reference}

\end{document}